\documentclass{JASSS}
	
\title{Synthetic Social Media Influence Experimentation via an Agentic Reinforcement Learning Large Language Model Bot}

\reviewcopy{false}

\author[1]{Bailu Jin}
\author[1]{Weisi Guo}
\affil[1]{Cranfield University, College Rd, Wharley End, Bedford MK43 0AL}

\email{weisi.guo@cranfield.ac.uk}

\usepackage{times}
\usepackage{helvet}
\usepackage{courier}
\usepackage{algorithm}
\usepackage{xcolor}
\usepackage{graphicx}
\usepackage{subcaption}
\usepackage{algpseudocode}
\usepackage{amsmath}
\usepackage{url}
\usepackage{natbib}
	\setcitestyle{authoryear,round,aysep={}}
	
\begin{document}
\maketitle 



\begin{abstract}
Understanding the dynamics of public opinion evolution on online social platforms is crucial for understanding influence mechanisms and the provenance of information.  Traditional influence analysis is typically divided into qualitative assessments of personal attributes (e.g., psychology of influence) and quantitative evaluations of influence power mechanisms (e.g., social network analysis). One challenge faced by researchers is the ethics of real-world experimentation and the lack of social influence data. In this study, we provide a novel simulated environment that combines agentic intelligence with Large Language Models (LLMs) to test topic-specific influence mechanisms ethically. Our framework contains agents that generate posts, form opinions on specific topics, and socially follow/unfollow each other based on the outcome of discussions. This simulation allows researchers to observe the evolution of how opinions form and how influence leaders emerge. Using our own framework, we design an opinion leader that utilizes Reinforcement Learning (RL) to adapt its linguistic interaction with the community to maximize its influence and followers over time. Our current findings reveal that constraining the action space and incorporating self-observation are key factors for achieving stable and consistent opinion leader generation for topic-specific influence. This demonstrates the simulation framework's capacity to create agents that can adapt to complex and unpredictable social dynamics. The work is important in an age of increasing online influence on social attitudes and emerging technologies.
\end{abstract}


\parano{}

\section{Introduction}

The emergence of online social media has revolutionized the way we share information and conduct debate. In order to answer key social research questions, we often have to create social influence simulation engines that involve modeling agents (how they behave and learn), how the agents socially network together, and how they disseminate information to each other.

Generally, Agent-Based Modeling (ABM) have proven effective in analyzing and explaining social phenomena by incorporating personal and environmental details. ABMs can use expert-driven models or data-driven learning functions to capture the nonlinear societal mechanisms and reveal large-scale societal emergent effects \cite{jackson2017agent}. Within agent modeling, reinforcement learning (RL) has the capability to mimic the human-like action selection. Using RL model, the target agent learn optimal strategies based on the interaction with social environment. These models have been developed to capture our understanding of a social environment that involves multiple agents interacting through language and physical encounters. The social observations that often interest us include: social consensus, clustering, and bi-polarity in shaping public opinion. ABMs also allow computation social science researchers to test hypothesis in more ethical environments without direct real world implications, and create useful data sets that are often lacking from the real world. A current limitation is that ABMs primarily function through mathematical representations and do not inherently integrate natural language behaviour. Emerging research, which we review in Section II will focus on developing Large Language Models (LLMs) that are controlled by RL agents to ingest and produce natural language interactions with other agents.

\subsection{Research Problem Statement and Aim}

There remain gaps between traditional small-scale psychological analysis of peer influence and the macroscopic analysis of large-scale social media influence. These gaps include the areas of intricacy, discontinuity, and heterogeneity in individual behaviors. Our research aim to bridge this gap between computational simulations and linguistic data by integrating RL-agentic intelligence and Large Language Models (LLMs). 

In our research, we have developed a simulated environment that replicates the dynamics of group discussions on a specific topic in an online social media context. This environment is structured with agents representing individual participant and follow relationships as links. Agents release topic-related posts to represent the current opinion and  the links will be updated due to opinion states. LLMs are used for post generation and opinion elicitation for agents. The generated posts are not only relevant to the topic but also reflect the complex influences present in a social media environment. The simulation result aligns with the polarity distribution of the actual dataset, indicating that our simulated environment reflects realistic discussion patterns.

\subsection{Novelty and Contributions}

In the paper, we specifically focus on training RL models on a target agent with the objective of maximizing its follower count in the simulated network as a proxy measure for opinion shaping. Our study explores various environment and observability settings to understand their impact on the agent's ability to achieve this goal. In our designed cases , the convergence in RL reward learning curves indicating RL achieved optimal solutions.

Here we summarize the major contributions of this paper:
\begin{itemize}
    \item  Developing a simulation framework on online social networks that integrate natural language interactions through LLMs.
    \item Integrating agents that can make decisions on how to improve its influence using reinforcement learning.
    \item Our research findings show that limiting the action space of the agent and incorporating self-observation are key factors for stable generation of opinion leaders.
\end{itemize}
The potential applications of this research are diverse across multiple domains. Influence models can allow us to better understand where misinformation has come from and why certain sources are more successful than others. This affects both policy makers as well as homeland security. For instance, social acceptance of emerging technologies like nanotechnologies and future 6G telecommunications is crucial for business and government (\cite{briguglio2021business}), policymakers could use the tool to simulate and evaluate the public's response to communication strategies. In business intelligence, marketing firms could leverage the model to predict the effectiveness of promotional strategies by simulating how opinions evolve in response to targeted campaigns.


\section{Related Work}

\subsection{Opinion Leader}
As discussed in Section 1, we first review existing work in ABM and how RL is a particular way for the agent to learn from data on achieving a desired social effect (e.g., opinion influence) in an environment. Here we review detailed works in ABMs, RL models, and language model interaction environments related to social opinion dynamics.

\subsection{Agent-Based Modelling (ABM)}

Agent-based modelling has been widely used by social researchers in the field of social simulation. In the modelling, there are generally two elements: (1) how agents interact with each other through the exchange of information, and (2) how agents process any information and adapt their behaviour and/or opinion of a topic. Compared to the other approaches, such as laboratory or filed experiments, ABMs apply the experimental control on a large-scale population to capture the nonlinear societal mechanisms and reveal large-scale societal emergence (\cite{jackson2017agent}). However, high control of ABMs leads to a low degree of external validity. To solve this problem, Betz presented a natural-language agent-based model of argumentation (ABMA) to simulate the argumentative opinion dynamics (\cite{betz2021natural}). The explicit stance used in traditional simulation of opinion dynamics can be represented by implicit argument from natural-language ABMAs. A number of recent works have used ABMs to integrate generative language models (GABMs) to create social networks. 

One approach to model realistic agent interactions is to simulate the life of agents in a sandbox and information exchange occurs through physical encounters \cite{park2023social}. Physical interactions aid the spread of information, and this has been expanded to model diverse demographics \cite{park2024}. To reduce the complexity of simulations, other models adopt a norm-diffusion model via LLMs \cite{Ghaffarzadegan24}. Our approach differs to these connection mechanisms in the sense that we do not prescribe a particular mechanism in how influence is spread, instead, we ask the RL agent to intelligently find the best strategy/policy to achieve influence spread by saying the right words and causing connections to be formed or removed based on consensus. This means it always has to adapt its output based on how the influence is being spread currently. Perhaps statistically this eventually leads to a diffusion-like mechanism, but this is beyond the scope of our scientific goals.

\begin{table}[ht]
\centering
\begin{tabular}{ccc}
\hline
Research Focus  & Data / Environment & Influence Mechanism \& Agent Design \\
\hline
Debate \& Arguments (\cite{betz2021natural}) & Verbal Simulation & Symbolic reason balance model \\
Spatial influence (\cite{pmlr-v97-jaques19a}) & Maze world & RL tasks \\
Opinion dynamics (\cite{R1}) & SBC Model & Deep RL \\
Generative diffusion (\cite{Ghaffarzadegan24}) & Simulation & Diffusion with LLM \\
Explicit dynamic graphs (\cite{10068693}) & OSN data sets & Graph-ODEs \& Neural Nets (\cite{10552434}) \\ 
Language informed tasks (\cite{Dimi24}) & Simulation & LLM driven Q-values in RL \\
Real-world experience (\cite{park2023social}) & Simulation world & Memory-stream \& Reflection driven LLM \\
Our proposed work   & OSN topic data sets & RL with LLM output \\
\hline
\end{tabular}
\caption{Review of Current Work and Positioning of Our Work. Our key contributions are: (i) topic specific real-data, and (ii) no assumed influence mechanism - we utilize RL-driven influence strategies.}
\label{Review}
\end{table}

\subsection{Agent Design using Reinforcement Learning (RL)}

As previously discussed, in RL agent design, the target agent learn optimal strategies based on the interaction with social environment. Originating from animal learning in psychology, RL has the capacity to mimic the human-like action selection. In 1989, \cite{watkins1992q} significantly advanced the field by integrating the theory of optimal control with temporal-difference (TD) learning, leading to the development of Q-learning. To address the curse of dimensionality in Q-learning, \cite{mnih2015human} combined deep learning with RL to create the deep Q-network (DQN), which notably surpassed professional human players in a range of 49 classic Atari games. We know from evidence that social conformity and influence is closely associated with RL and hence our design is motivated by evidence from \cite{review_cell}, and used in advertising to increase social influence modeled via Stochastic Bounded Confidence Model (SBCM) \cite{R1}.

Recent advances have extended the application of RL in various domains in natural sciences, social sciences, and engineering (\cite{silver2016mastering}, \cite{social_robot} and \cite{luo2017adaptive}). In the work of \cite{sert2020segregation}, authors show that the combination of RL and ABM can create an artificial environment for policy makers to observe potential and existing behaviors associated with the rules of interactions and rewards. These RL agents often involve using graph embeddings as a way to articulate the importance of network centrality in social influence mechanisms \cite{review_social_network_influence}, such as diffusion and consensus through synchronization \cite{10552434}. A challenge in current literature is that often the opinion or influence is mathematically modeled in a general way. For example, \cite{10068693, review_opinion} projected online social network (OSN) data through NLP dimension down-scaling onto ordinary differential equation (ODE) dynamic social graphs. These approaches \cite{Shepherd_Goldsmith_2020} lack a more naturalistic interface that is compatible with how humans interact and as such also lack topic-sensitivity. RL agents have also been actively applied to social robotics to exert influence \cite{pmlr-v139-ndousse21a}, but the interplay with LLMs to enable more naturalistic behaviour and human interaction has been a growing recent research topic. There are also spatial applications of RL driven influence as shown in \cite{pmlr-v97-jaques19a}, where robots use their spatial navigation and task solving ability to create influence on others. The integration of LLM into RL has also recently advanced, for example using LLM knowledge to inform the attention space (e.g., Q-values) of RL agents \cite{Dimi24}.

\subsection{Large Language Model}

LLM such as GPT has been tested for multiple natural language processing tasks and proved to be effective without gradient updates or fine-tuning \cite{brown2020language}. Since GPT-3 was trained on large web corpora which include social media behavior, GPT-3 can predict user’s future behaviours, responses, or action plans with some accuracy. GPT was able to give human-like response to replace the expensive and time-consuming user studies (\cite{sekulic2022evaluating}). Although the generated synthetic data was proved to exhibit less language variability, it may lead to better classification results when few data is available and resources are limited (\cite{meyer2022we}). Recent study has use Large Language Models to simulate the online social interactions, leading to improved dialogue and outcomes through debate \cite{R21}. LLMs are used to predict the popularity on the online social media (\cite{park2022social}), the distribution of votes for elections, the sentiment of news, and the voting of justices (\cite{hamilton2023blind}). In the work of \cite{park2022social}, authors presented a social simulacra tool, SimReddit, which is a prototyping technique generating social behaviours using large language models and matching to complexity analysis \cite{Lu24Complex}. With the description of a community's design, the social simulacra produce appropriate simulated behavior. Game theoretic frameworks have also been explored using LLMs \cite{Lore24}.

\begin{figure*}[ht]
    \centering    
    \includegraphics[width=0.9\linewidth]{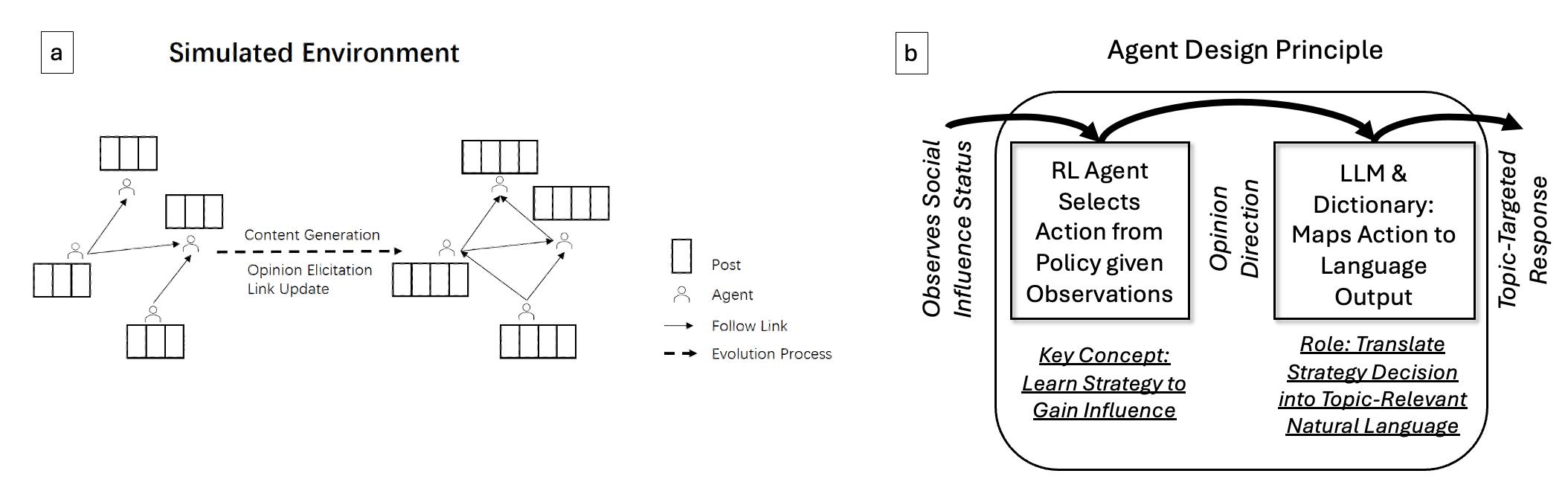}
    \caption{(a) Simulated Environment Social Network: Participants within this environment, referred to as 'agents', are interconnected through the 'follow' relationship between each other. Agents interact through 'Posts' which is a small natural language message, formatted like a Tweet. Social influence evolution process shows the development of the whole network from $t$ to $t+1$. The process can be segmented into three part: Post Generation, Opinion Elicitation and Link (follow) Update. This process is how we track and measure influence. (b) Agent Design Principle: The agent is comprised of 2 building blocks - (i) the RL model that formulates how best to strategically gain influence in the long term, and (ii) the LLM that maps the strategy decisions from RL to natural language tweets for the topic specified.}
    \label{fig:SimulatedEnvironment}
\end{figure*}


\section{Simulated Social Environment}

One challenge faced by researchers is the ethics of real-world experimentation and the lack of social influence data. The purpose of the simulated environment is to better understand how online social members influence each other through dialogue and follow/unfollow mechanisms. Our environment contains agents that generate posts, form opinions on topics, and socially follow each other based on the outcome of discussions. This simulation allows researchers to observe the evolution of how opinions form and how influence leaders emerge. In the following section, we detail how an agent can employ Reinforcement Learning (RL) to optimise its interaction with the community with the purpose of maximizing its influence and following. We then detail the results and findings of our computation experiments. 

\subsection{Environment}
In this section, we introduce the development of a simulated environment that replicates the dynamics of group discussions on online social media platforms. Discussions in this environment revolve around the topic, defined by posts that represent divergent viewpoints within the conversation.

The simulated environment system graph is shown in Figure \ref{fig:SimulatedEnvironment}a. We will explain each part of the system step by step.

Within this simulated setting, a post is a small natural language message, formatting like Tweet. These posts are contributing to the whole discussion, each either supporting or opposing the central topic.

For instance, in our experiment, we initiate discussions using the topic:'Society with no gender'. Posts such as 'This incorrectly supposes that some current biological sex traits aren't functional. I think a society without gender is a really bad idea.' are categorized as 'con' post to the topic. Conversely, posts like 'The absence of gender constructions would enable people to enjoy a better overall quality of life.' support the topic, as a 'pro' post.

Participants within this environment, referred to as 'agents', are interconnected through the 'follow' relationship between each other. Follow relationship is unidirectional. Each agent generates a post $\text{POST}^t_i$ at time t, influenced by their existing follow links and previous posts. Consequently, if $\text{agent}_i$ follows $\text{agent}_j$ at time $t$, then $\text{POST}^t_j$ potentially impacts $\text{POST}^{t+1}_i$.

At any given time$t$, $\text{POST}^t_i$ fully determines the agent $i$'s opinion $\text{OPIN}^t_i$. The network of follow links between agents is then dynamically updated in response to these evolving $\text{OPIN}^{t+1}$.  

The process can be segmented into three part: Post Generation, Opinion Elicitation and Link Update, as outlined in the following Algorithm\ref{alg:envir}.

\begin{algorithm}
\caption{Simulated environment algorithm}\label{alg:envir}
\begin{algorithmic}
\For{\(t \in [1 \ldots t_{\text{max}}]\)}
    \For{\(i \in \text{AGENTS}\)}
        \State locate the follow link of agent \(i\) 
        \State generate the post of agent \(i\) (\( \rightarrow \text{POST}_{i}^{t}\));
        \State elicit the opinion of agent \(i\) (\( \rightarrow \text{OPIN}_{i}^{t}\));
    \EndFor
\State update the follow link of AGENTS;
\EndFor
\end{algorithmic}
\end{algorithm}

\subsection{Post Generation}

This section explores the mechanics of LLMs in content generation technology. These models predict the next-word in a sequence with a probabilistic approach given the prompt input.

The objective of the post generation stage is to generate the possible post of agent at the given time. This process incorporates the agent's previous posts and those from their followees. The post is generated by the LLM based on instructions received from RL model action space (see Figure \ref{fig:SimulatedEnvironment}b). As such, the LLM is not involved in action selection of the RL, rather it is a plug-in that translates opinion action space into natural language for dialogue

The generated post must be concise a short sentence formatting like Tweet. Therefore we use the natural language description to describe the follow relationship, choose the format of web API including a designated stop sign as ('\}') to ensure the output aligns with the expected structure.  

Consider the following example: if $\text{agent}_i$ follows $\text{agent}_j$, the prompt for generating  $\text{POST}^{t+1}_i$ would be:

\begin{quote}
    $\text{agent}_i$  posted tweets about 'a society without gender': $\text{POST}^{t}_i$ 
    
    $\text{agent}_i$ saw the following tweets about 'a society without gender' on homepage\{'data':\{'user': $\text{agent}_j$,'text': $\text{POST}^{t}_j$ \}\}. "
    
    $\text{agent}_i$ shared the following tweet on this topic:\{'data: \{'user': $\text{agent}_i$,'text':'"
\end{quote}

The first paragraph of the prompt describe the topic and refer the previous posts from $\text{agent}_i$. The second paragraph then suggests $\text{agent}_i$ follows $\text{agent}_j$, so the post of $\text{agent}_j$ will potentially impact the next post of $\text{agent}_i$. If $\text{agent}_i$ follows multiple agents in the group, the prompt will include more content that $\text{agent}_i$ can read from the homepage. The third paragraph ensure $\text{agent}_i$ still discuss the same topic and generation of  tweet-like post. The name of agents, allowing for the observation of interaction between them, such as agreements or disagreements on the topic. An example of a generated post, demonstrating such an interaction, is as follow:
\begin{quote}
    "I agree with both Grace Lee and Chloe Kim. Gender is an important aspect of our identity and should not be disregarded. However, everyone should also have the freedom to express themselves and live authentically. Instead of erasing gender, we should focus on promoting equality and challenging harmful gender stereotypes. Let's create a society where individuals are free to express their gender identity without fear of discrimination or prejudice. \#GenderEquality \#BeYourself"
\end{quote}

This response reflects incorporating with other agents and presenting a balanced perspective. It demonstrates that the generation of posts is not only topic-relevant but also reflective of the influences in a social media environment.

Moving to text generation process, key stages include 'encoding', where the model assimilates and interprets the input text, identifying contextual patterns, and 'decoding', where it generates new text from this analysis.

Crucial to the flexibility of these models in the decoding phase are key parameters such as 'temperature' and 'top\_p'. Temperature controls the randomness in the prediction, influencing creativity and coherence in the generated text. 'top\_p', controls the model to focus on a subset of the most probable next words, ensuring relevance and precision in the output. These two parameters balance the generated content between novelty and fidelity to the input context.

\begin{table}[ht]
\centering
\begin{tabular}{c c c}
\hline
 & temperature & top-p \\
 \hline
Narrow & 0.1 & 0.5 \\ 
Creative & 1.4 & 0.95 \\ 
\hline
\end{tabular}
\caption{GPT Decoding Setting}
\label{gptSetting}
\end{table}

Table.\ref{gptSetting} shows the our setting for narrow case and creative case. Low temperature and low top-p lead to conservative, and narrow\- minded agents who're sticking with the most obvious options when generating a post. The creative setting characterize agents who are much willing to take surprising turns.

\subsection{Opinion Elicitation}

The opinion elicitation process gives the probability that the agent expects a pro-claim rather than a con-claim given the perspective using perplexity calculation. The method we used is introduced in the work of Betz \cite{betz2021natural}, where we follow its steps but with a different purpose. Betz aimed to discover whether introducing new arguments will disrupt the dynamics of collective deliberation, whereas our paper aims to improve the number of followers/social influence by introducing a RL-LLM agentic intelligence to our agent.

Perplexity effectively measures how well a probabilistic model predicts a sample, serving as an indicator of the model's predictive capacity across the entire set of tokens within a corpus. In the context of natural language processing, perplexity is defined as the exponential of the average negative log-likelihood of a given sequence of tokens. Consider a tokenized sequence represented as $S = (w_0,w_1,...,w_m)$. The perplexity of $S$ denoted as $\text{PPL}(S)$, is computed as:

\begin{equation}
    \text{PPL}(S) = \exp\left(-\frac{1}{m} \sum_{k=0}^{m} \log p_{\theta}(w_k | w_{<k})\right)
    \label{ppl}
\end{equation}

In equation (\ref{ppl}), $\log p_{\theta}(w_k | w_{<k})$  represents the logarithm of the likelihood of the $k$th token in the sequence, given all preceding tokens(denoted as $w_{<k}$) as per the probabilistic model in use. 

When GPT processes a sentence, it calculates the logarithm of the probability of each word given the previous words in the sentence, which is $\log p_{\theta}(w_k | w_{<k})$ in the equation.

\begin{equation}
    \text{PPL}_{\text{CON}}(\text{POST}_t^i) = \text{PPL}(S_\text{CON}|\text{POST}_t^i))
    \label{con}
\end{equation}

\begin{equation}
    \text{PPL}_{\text{PRO}}(\text{POST}_t^i) = \text{PPL}(S_\text{PRO}|\text{POST}_t^i))
    \label{pro}
\end{equation}

Equation (\ref{con}) shows how to compute ${\text{PPL}_{\text{CON}}(\text{POST}_t^i)}$. Given a predefined sentence $S_\text{CON}$ like "society without gender is a really bad idea", the conditional perplexity shows the inverse probability of the output of the language model. Similarly, ${\text{PPL}_\text{PRO}(\text{POST}_t^i)}$ is computed using the predefined sentence "Society without gender is a really good idea". The low perplexity result suggests that the language model predicts this particular sentence with a high level of certainty.

\begin{equation}
    \text{OPIN}_t^i = \frac{\text{PPL}_{\text{CON}}(\text{POST}_t^i)}{\text{PPL}_{\text{PRO}}(\text{POST}_t^i) + \text{PPL}_{\text{CON}}(\text{POST}_t^i)}.
    \label{opin}
\end{equation}

Based on the conditional perplexity scores for pro and con sentences, which indicate the confidence of the language model on each side of perspective, we compute the overall opinion score. In the equation (\ref{opin}) we give the definition of $\text{OPIN}_t^i$. The outcome is a float number in range $[0,1]$ which represent the polarity where 0 represents con and 1 represent pro. We will then categorise the number into five polarity categories: 'strong con', 'con', 'neutral', 'pro' and 'strong pro'.

\subsection{Link Update}

To establish the ‘Following’ relationship between users, we adopt the premise that individuals inclined to follow users who express similar opinions, as suggested by \cite{zhou2009finding}. Once we have the categorised polarity of each tweet, we then hypothesise that individuals with the same polarity category hold similar opinions.

Therefore, we explore two link methods: a Follow Dynamics and a Follow-Unfollow Dynamics. In both cases, if two users consistently align in the same opinion category over a few consecutive steps, they are considered likely to follow each other, with a follow probability. In Follow-Unfollow Dynamics, the exist follow link may also break with an unfollow probability, if the opinions of two people differ during the consecutive steps. Details of follow link update and unfollow link update are given in Algorithm \ref{alg:update_link} and Algorithm \ref{alg:un_update_link}. 

\begin{algorithm}
\caption{Follow link update algorithm}\label{alg:update_link}
\begin{algorithmic}

\For{each agent $i$}
    \For{each agent $k \neq i$ not already followed by $i$}
        \For{$t \in [1, \text{num\_same}]$}
            \If{ $\text{OPIN}_{i}^{t}$ and 
            $\text{OPIN}_{k}^{t}$ in the same opinion category}
                \State \textbf{continue}
            \Else
                \State $\text{match} \gets \text{false}$
                \State \textbf{break}
            \EndIf
        \EndFor
        \If{loop completed without break}
            \If{random number $<$ \text{follow link rate threshold}}
            \State $\text{match} \gets \text{true}$
            \State Add link from $i$ to $k$ in $\text{new follow dict}$
            \EndIf
        \EndIf
    \EndFor
\EndFor
\end{algorithmic}
\end{algorithm}

\begin{algorithm}
\caption{Unfollow link update algorithm}\label{alg:un_update_link}
\begin{algorithmic}

\For{each agent $i$}
    \For{each agent $k \neq i$ already followed by $i$}
        \For{$t \in [1, \text{num\_same}]$}
            \If{ $\text{OPIN}_{i}^{t}$ and 
            $\text{OPIN}_{k}^{t}$ not in the same opinion category}
                \If{random number $<$ \text{unfollow link rate threshold}}
                \State $\text{match} \gets \text{false}$
                \State Delete link from $i$ to $k$ in $\text{new follow dict}$
                \State \textbf{break}
                \EndIf
            \EndIf
        \EndFor

    \EndFor
\EndFor
\end{algorithmic}
\end{algorithm}

\begin{figure*}[ht]
    \centering    \includegraphics[width=0.9\linewidth]{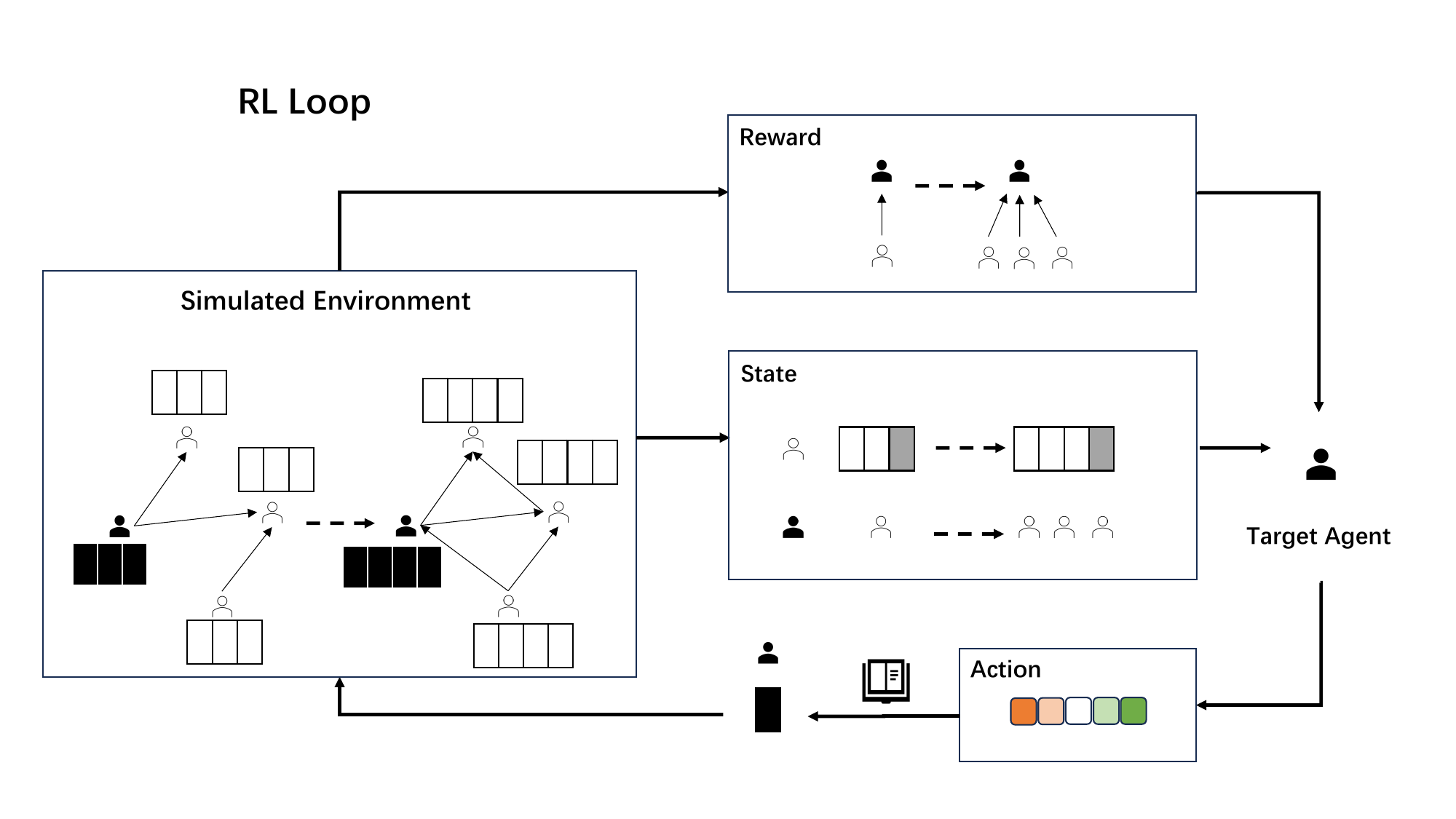}
    \caption{RL Loop. Reward: change of number of followers at each time step. State: 1)the current opinion states of the initial accounts followed by the target agent, 2)the number of followers the target agent currently has. Action: pre-defined five opinion polarity categories.}
    \label{fig:RLsystem}
\end{figure*}

\section{Agent Design: Reinforcement Learning}

In the previous part, we have introduced the whole simulated environment. In this part, we detail the reinforcement learning setting employed for the model. 

\subsection{Model Design}

We set a target agent in this group network without any initial followers. Then the target agent can actively choose current opinion state and post related posts. The objective is try to let the target agent get more followers within the determined time step.

The reinforcement learning (RL) model utilized in this framework is based on the q-learning algorithm, a widely used method in machine learning for environments characterized by uncertainty. This technique is used for developing optimal strategies through learning from interactions within such environments.

The RL Loop structure is shown in Figure \ref{fig:RLsystem}. The target agent interacts with the previous introduced simulated environment. At each time step, the agent observes the state of the environment, selects an action, receives a reward from the environment, leading to a new environmental state.

\subsubsection{Reward}
The reward mechanism for this model is the change of number of followers at each time step, encouraging strategies that enhance follower acquisition.

\subsubsection{State}
To determine what a target agent can observe, we assume that the agent has visibility over the accounts it follows, as well as its own state. The observed state is a matrix which comprises two elements: 1) the current opinion states of the accounts followed by the target agent, 2) the number of followers the target agent currently has.

\subsubsection{Action}
At the action stage, target agents participate in the discussion by selecting a sentence to express their opinion. The available action space consists of five predefined opinion polarity categories: strong con, con, neutral, pro, and strong pro. The LLM is not involved in action selection of the RL as shown in Figure 1b, rather it is a plug-in that translates opinion action space into natural language for dialogue.

In our experiment, we use the actual posts to generate an action dictionary. The dictionary maps actual posts to opinion categories. After selecting an opinion category based on the RL algorithm, the target agent randomly chooses a tweet from the corresponding category.

\subsection{Reinforcement Learning Algorithm}
\begin{algorithm}
\caption{Reinforcement Learning Algorithm}\label{alg:RL}
\begin{algorithmic}
\State Initialize Q-values $Q(s, a)$ arbitrarily for all state-action pairs $(s, a)$.
\For{each episode}
    \State initialize state $s$
    \For{\(t \in [1 \ldots t_{\text{max}}]\)}
        \State target agent choose action $a$ from state $s$ using policy derived from Q.
        \For{\(i \in \text{AGENTS} \) and $i$ is not target agent}
            \State generate the post of agent \(i\) (\( \rightarrow \text{POST}_{i}^{t}\));
            \State elicit the opinion of agent \(i\) (\( \rightarrow \text{OPIN}_{i}^{t}\));
        \EndFor
    \State update the follow link of AGENTS;
    \State observe reward $r$, and new state $s'$;
    \State Update Q-value for $(s, a)$;
    \State Update state $s \leftarrow s'$;
    \EndFor
\EndFor
\end{algorithmic}
\end{algorithm}

Referencing Algorithm \ref{alg:RL}, the process begins with the initialization of Q-values for all state-action pairs. In each episode, the state is initialized, and the agent embarks on a series of time steps. During each time step, the target agent selects an action from its current state, based on a policy derived from the Q-values. Alongside each agent in the system generates posts and forms opinions. The algorithm then updates the follow links among agents and processes the resulting reward (followers) and new state. This leads to an update in the Q-value for the chosen action, and the agent's state is updated accordingly.

The Q-value update is represented by the formula:

$$    Q(s, a) \leftarrow Q(s, a) + \alpha \left[ r + \gamma \max_{a'} Q(s', a') - Q(s, a) \right]$$

In this formula, \( \alpha \) represents the learning rate, \( \gamma \) the discount factor, \( r \) the reward (followers), and \( Q(s', a') \) the maximum estimated future reward. This update rule is key to the learning process, as it adjusts the Q-values based on the reward received and the potential future rewards, leading to more informed and strategic decision-making by the agent in subsequent steps. The details of the RL design and parameter tuning can be found in Table\ref{RLSetting}. We use Gender case and Drug case, set number of episodes 500, max of steps per episode as 5. Other opinion experiment specific parameters are shown in Appendix-D.

\begin{table}[ht]
\centering
\begin{tabular}{cc}
\hline
Agentic RL Design Parameter / Function  & Design Approach / Value \\
\hline
RL type             & Q-learning \\
reward function     & the number of followers \\
observation states  & opinion states of other agents followed, number of followers, self-observation \\
action space        & 5 levels of opinions \\
data case studies   & (1) gender debate, (2) drug-use debate \\
\hline
   learning rate    & 0.01 \\
    discount rate   & 0.99 \\
    epsilon         & 0.005 \\
    exploration rate & 0.01 \\
    max. steps per episode & 5 \\
    no. episodes & 500 \\
\hline
\end{tabular}
\caption{Q-learning RL Design and Training Parameters}
\label{RLSetting}
\end{table}

\section{Experiment}

\subsection{Simulated Environment Initialisation}

The initiation of our simulated environment begins with the creation of initial posts, for which we use the dataset released by \cite{betz2021natural}. The dataset is crawled from debating platform kialo.com, comprises various posts structured around specific topics. For our simulation, we select posts related to the topic 'Gender' and topic 'Drug'. 
The dataset includes 486 posts under 'Gender' topic and 660 posts under 'Drug' topic, with each posts labeled with one type 'pro' or 'con'. The length of each post is less than 70 words. For example, under 'Gender' topic, a post such as 'The absence of gender constructions would enable people to enjoy a better overall quality of life.' is labeled as type 'pro'.

\begin{figure*}[t]
    \centering
    \begin{subfigure}{0.30\textwidth}
        \centering
        \includegraphics[width=\linewidth]{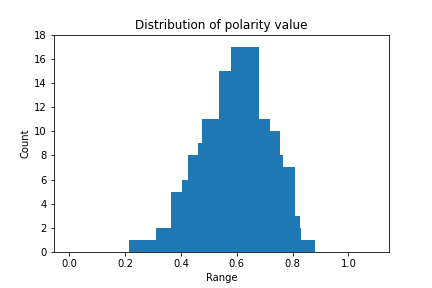} 
        \caption{Gender}
        \label{fig:distribution1}
    \end{subfigure}
    \begin{subfigure}{0.30\textwidth}
        \centering
        \includegraphics[width=\linewidth]{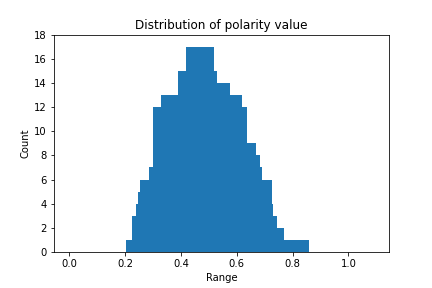} 
        \caption{Drug}
        \label{fig:distribution2}
    \end{subfigure}
    \caption{Distribution of polarity value.}
    \label{fig:distribution}
\end{figure*}

Employing the opinion elicitation process, we analyze the datasets to determine the polarity distribution. As shown in Figure \ref{fig:distribution}, the distributions of polarity value in both cases cluster around the middle range. This distribution is indicative of the varying degrees of support or opposition to the topic within this dataset.

Based on the polarity distributions, we categorize the posts into distinct opinion groups while ensuring an even distribution across these categories. The specific range settings are detailed in Appendix D, and the categorization results are presented in Table \ref{category} and Table \ref{Dcategory}.

\subsection{Simulated Environment Analysis}

To examine the opinion evolution process within our simulated environment, we run two 50-step simulations using both Narrow and Creative agent setting using Gender topic. These simulations aim to demonstrate how different agent settings can influence the dynamics of the discussion.

\begin{figure*}[ht]
    \centering    \includegraphics[width=1\linewidth]{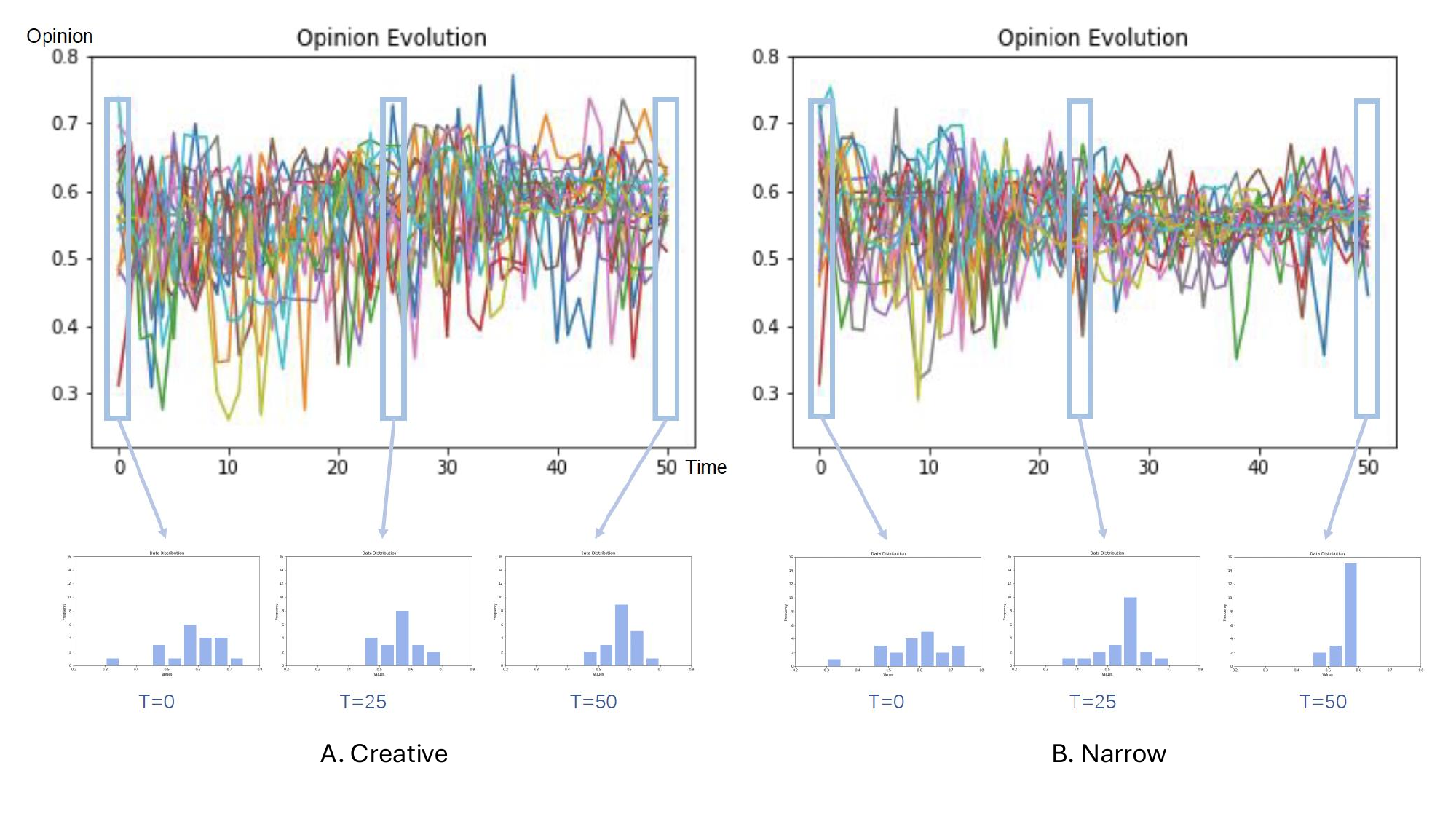}
    \caption{Opinion Evolution Visualization on Narrow setting and Creative setting}
    \label{OpinionEvolution}
\end{figure*}

\begin{figure*}[ht]
    \centering    \includegraphics[width=0.5\linewidth]{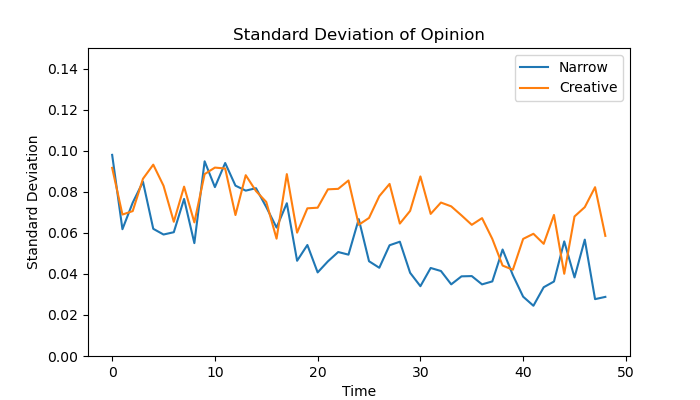}
    \caption{Standard Deviation of Opinion on  Narrow setting and Creative setting}
    \label{OpinionSTD}
\end{figure*}
    
    
    
 
Figure \ref{OpinionEvolution} illustrates the dynamics of opinion evolution under the influence of two simulation configurations: the Narrow setting and the Creative setting. Table \ref{gptSetting} provides the specific Temperature and Top-p settings used for the opinion dynamics simulation.

Opinion evolution visualizations show the trajectory of opinion polarity score for 20 agents over 50 steps on Narrow setting and Creative Setting.  Initially, the agents' opinions range from 0.3 to 0.8. In the Narrow setting, we observe that opinion density increases over time, showing a tendency toward opinion convergence. Conversely, in the Creative setting, the opinion density remains more dispersed and chaotic throughout the entire time scale, with less apparent convergence.

To provide a more detailed view of opinion dynamics, opinion distributions at Time 0, 25, and 50 are presented. In the Creative setting, despite the initial high degree of opinion diversity, after 25 steps, some grouping starts to occur, but full convergence is not reached even by step 50. The discussion remains dynamic, with polarity scores fluctuating in the noisy range between 0.4 and 0.7. On the other hand, in the Narrow setting, while there is significant opinion diversity in the first 25 steps, this diversity gradually decreases, and by step 50, opinions cluster in the 0.45 to 0.6 range, indicating a clearer sign of convergence.

In both scenarios, the distribution of the final polarity scores shows a slight skew towards higher values across the time steps. 

Figure \ref{OpinionSTD} plots the standard deviation of opinion over time in both Narrow and Creative settings. The graph shows that in the Narrow setting, the opinions form a more centralized distribution by step 50, indicating stronger clustering and a more definitive convergence of opinions, as compared to the Creative setting, which remains more dispersed throughout the simulation.

\begin{figure*}[t]
    \centering
    \begin{subfigure}{0.48\textwidth}
        \centering
        \includegraphics[width=\linewidth]{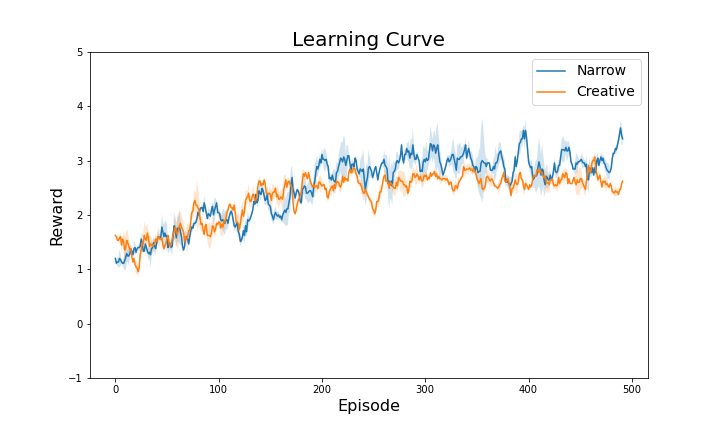} 
        \caption{Follow Dynamics, Part-Observable}
        \label{fig:sub1}
    \end{subfigure}
    \begin{subfigure}{0.48\textwidth}
        \centering
        \includegraphics[width=\linewidth]{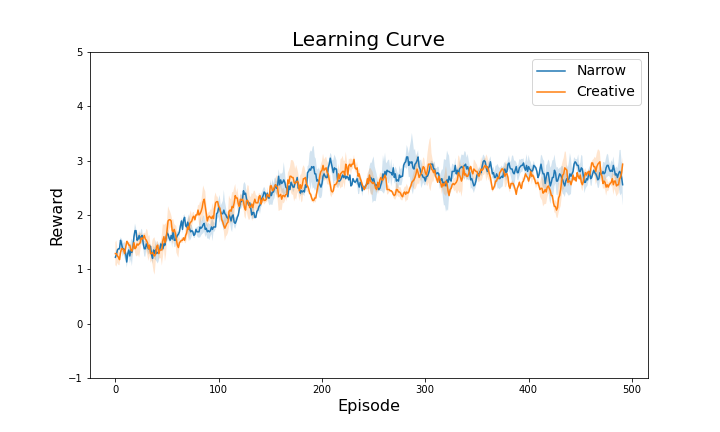} 
        \caption{Follow Dynamics, Full-Observable}
        \label{fig:sub2}
    \end{subfigure}
    \begin{subfigure}{0.48\textwidth}
        \centering
        \includegraphics[width=\linewidth]{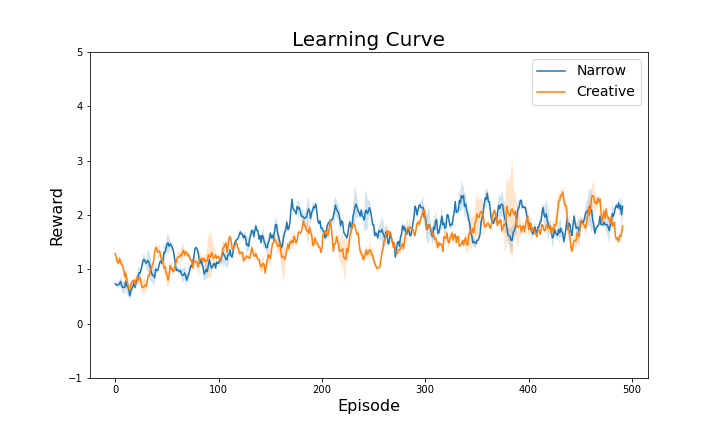} 
        \caption{Follow-Unfollow Dynamics, Part-Observable}
        \label{fig:sub3}
    \end{subfigure}
    \begin{subfigure}{0.48\textwidth}
        \centering
        \includegraphics[width=\linewidth]{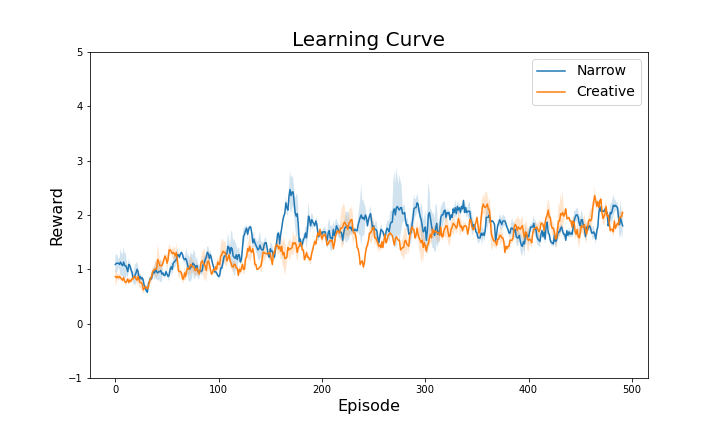} 
        \caption{Follow-Unfollow Dynamics, Full-Observable}
        \label{fig:sub4}
    \end{subfigure}
    \caption{Learning Curves of RL in Gender case, with different language model settings, link update methods, and observability settings.}
    \label{fig:RLResult1}
\end{figure*}

\begin{figure*}[t]
    \centering
    \begin{subfigure}{0.48\textwidth}
        \centering
        \includegraphics[width=\linewidth]{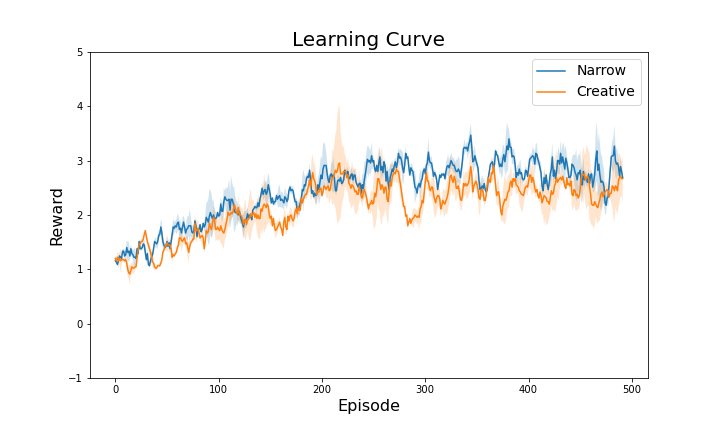} 
        \caption{Follow Dynamics, Part-Observable}
        \label{fig:sub1}
    \end{subfigure}
    \begin{subfigure}{0.48\textwidth}
        \centering
        \includegraphics[width=\linewidth]{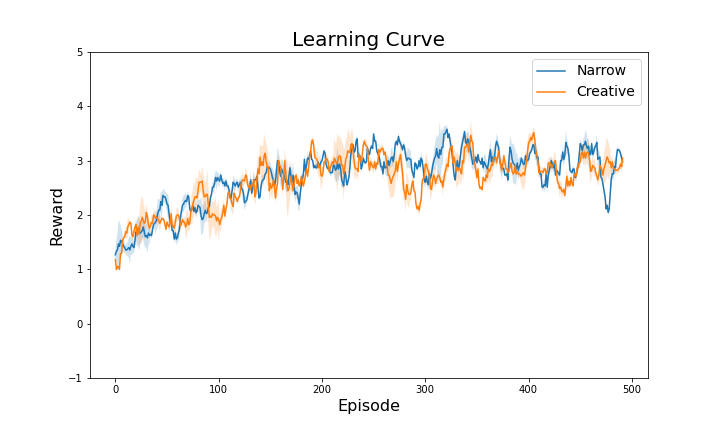} 
        \caption{Follow Dynamics, Full-Observable}
        \label{fig:sub2}
    \end{subfigure}
    \begin{subfigure}{0.48\textwidth}
        \centering
        \includegraphics[width=\linewidth]{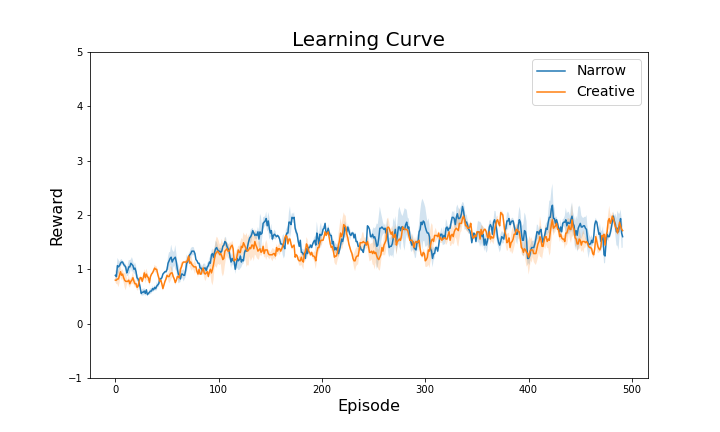} 
        \caption{Follow-Unfollow Dynamics, Part-Observable}
        \label{fig:sub3}
    \end{subfigure}
    \begin{subfigure}{0.48\textwidth}
        \centering
        \includegraphics[width=\linewidth]{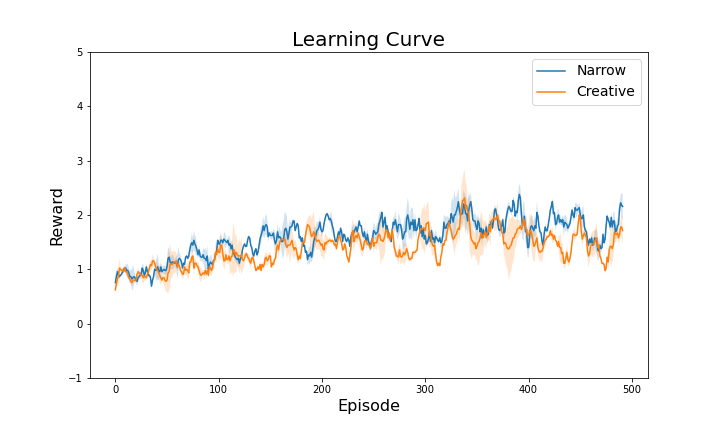} 
        \caption{Follow-Unfollow Dynamics, Full-Observable}
        \label{fig:sub4}
    \end{subfigure}
    \caption{Learning Curves of RL in Drug case, with different language model settings, link update methods, and observability settings.}
    \label{fig:RLResult2}
\end{figure*}

\subsection{Reinforcement Learning Setting}

In this section, we explore the application of RL within our simulated environment. The primary research question is: \textit{Can RL algorithms identify an optimal strategy that enables the target agent to maximize its follower count within a predetermined number of steps?}

By testing different settings, we found that limiting the action space of the RL agent and incorporating self-observation are key factors for stable RL result. We found that actions when restricted by using only statements in the dictionary, and not the LLM, significantly improved convergence in training due to the narrow topic of our experiments (e.g., gender rights). 

By restricting the action space, we reduce the complexity of the decision-making process (e.g., number of actions in a policy), enabling the RL agent to focus on a more manageable set of actions. This helps prevent the model from exploring either irrelevant or suboptimal actions within a policy that could introduce noise and instability into the learning process. Additionally, a smaller, well-defined action space allows for faster convergence, as the agent can more efficiently learn the consequences of its actions within the environment. 

Incorporating self-observation further enhances stability by allowing the agent to continuously assess its own state and progress. This self-awareness helps the agent make more informed decisions, as it can adjust its behavior based on how past actions have affected its follower count and other relevant metrics. Self-observation also promotes consistency, as the agent is better equipped to identify patterns in the environment and avoid erratic or random behaviors. Together, limiting the action space and incorporating self-observation contribute to generating more stable and reliable outcomes, ensuring that the RL agent can effectively optimize its strategy within the defined constraints of the simulation.

To investigate stable RL results, we conducted tests across eight scenarios.
\begin{enumerate}
    \item \textbf{Language Model Settings}:
    \begin{itemize}
        \item We employed two types of settings for the language models of agents, as delineated in Table \ref{gptSetting}: \textit{Narrow} and \textit{Creative}. These settings impact the nature of posts generated by the agents.
    \end{itemize}
    
    \item \textbf{Link Update Methods}:
    \begin{itemize}
        \item \textit{Follow Dynamics}: In this case, the follow rate is 0.8, and the unfollow rate is 0. This setting allows us to observe the growth of the target agent's network under a scenario where new follow links are relatively likely to form.
        \item \textit{Follow-Unfollow Dynamics}: In this more dynamic setting, the follow rate is still 0.8, and the unfollow rate is set as 0.5. This case allowing existing follow links to break, thus adding a layer of complexity to the network dynamics.
    \end{itemize}
    
    \item \textbf{Observability Settings}:
    \begin{itemize}
        \item \textit{Part-Observable Case}: In this scenario, the target agent initially follows only one other agent.
        \item \textit{Full-Observable Case}: Contrasting the Part-Observable setting, in this case, the target agent begins by following every agent in the network.
    \end{itemize}
\end{enumerate} 

We use Gender case and Drug case, set number of episodes 500, max of steps per episode as 5. Other parameters are shown in Appendix-Parameters.

\subsection{Reinforcement Learning Results}

This subsection presents the outcomes of our RL experiments conducted across the eight scenarios. Figure \ref{fig:RLResult1} illustrates the results for the Gender scenario and Figure \ref{fig:RLResult2} shows the results for the Drug scenario. In both figures, the colored lines represent the average reward (followers) of five repeated experiments, and the shaded areas indicate the variance in the results. The followers gained (reward) rises from 1 to 2 or 3 in different cases in a total agent population of 20. That is to say, we see a gain of 10\% of the total available follower population.

Regarding a baseline, if there was no intelligence from RL and we just used initial conditions, it would be a statistical flat line with no followers gained over time.  

In all scenarios, we observe convergence in the reward learning curves (followers), indicating that the RL algorithm successfully identified optimal solutions within the given settings. Both the Gender and Drug cases exhibit similar learning curve patterns under different conditions, highlighting consistent performance across varying contexts.

One notable observation is that all Follow-Unfollow Dynamics cases exhibited greater variance in their learning curves compared to the Follow Dynamics counterparts under same setting. This implies that more complex scenarios pose greater challenges in reaching optimal results.

Cases with Full-Observable settings demonstrates more stable learning curve than the part-observable ones. This suggests that target agent with complete network visibility are likely to learn strategies more effectively and with greater stability than those with limited visibility.

The Narrow and Creative settings, when applied to the Partially-Observable cases, presented distinct learning curves. The Narrow setting consistently outperformed the Creative setting, as evidenced by higher reward values. However, in the Full-Observable scenarios, the learning curves for both Narrow and Creative settings align closely. A possible explanation is that, in Part-Observable scenarios, the unpredictable nature of agents in the Creative setting, leading to lower performance compared to the Narrow setting. Conversely, in Full-Observable scenarios, the agents are better able to adapt to and anticipate the Creative setting's variable patterns.

In summary, these results highlight the impact of observability and complexity in reinforcement learning environments, with Full-Observable settings and simpler dynamics facilitating more stable and efficient learning outcomes. The Narrow setting's predictable nature resulted in higher performance than the unpredictable Creative setting. However, when observability was not a constraint, as in the Full-Observable scenarios, both settings showed similar performance. This suggests that with full visibility, agents can effectively adapt to even unpredictable dynamics.

\section{Conclusions and Future Work}

Understanding the dynamics of public opinion evolution on online social platforms is crucial for understanding influence mechanisms and the provenance of information. Traditional influence analysis is typically divided into qualitative assessments of personal attributes (e.g., psychology of influence) and quantitative evaluations of influence power mechanisms (e.g., social network analysis). 

Many of the analysis approaches assume how influence is spread (e.g., diffusion \cite{Ghaffarzadegan24}, opinion ODE \cite{10068693}, or stochastic bounded confidence models \cite{R1}), but these are not topic-specific and lack dynamic adaptation at the individual scale. Our approach uses reinforcement learning (RL) to strategically decide what actions to take in conversations and an LLM to translate the actions into natural language. This enables us to better modulate our conversations to be natural and topic-specific, whilst maximizing the individual agent's long-term influence upon peers. Whilst there is emerging work on integrating LLMs and RL agents \cite{Dimi24} and large-scale simulations of agent behaviour \cite{park2024}, we believe at time of writing this is an important contribution.

In terms of academic contribution, our work also enable researchers to ethically test new agents and create fine-scale experimentation on opinion dynamics for diverse topics.

Our current findings reveal the following: (1) RL can achieve a gain of 10\% of the total available follower population, and (2) convergence can be better achieved by limiting the action space and incorporating self-observation. This demonstrates the simulation framework's capacity to create agents that can adapt to complex and unpredictable social dynamics. The potential application would be providing valuable insights for digital marketing strategists and policymakers in the dynamics of online opinion formation and influence. However, we acknowledge that one potential negative social impacts could be the misuse of the simulated environment in manipulating public opinion. If used unethically, the technology may craft highly persuasive and targeted misinformation campaigns on social media platforms. Conversely, this technology can also be instrumental in detecting malicious activities in social networks.

One limitation of our research is the inherent unpredictability of actual social media discussions. Social dynamics are complex and unpredictable, and our rule-based simulated environment cannot fully replicate. Although the pre-trained RL algorithm demonstrates promising performs in the simulated environment, it presents one version of the actual possible outcome. Despite this, our research provides valuable insights into the mechanics of social interaction and opinion dynamics in online groups.

Our future work would be directed towards enhancing the realism and variability of our simulated social media environment. This may involve incorporating more sophisticated interaction rules to better mirror the real-world social media dynamics. To achieve this, first we plan to discover how different character profiles of agents perform and interact within the network. This exploration aims to provide a deeper understanding of the varied behaviors and their impacts on the overall dynamics of social media interactions. We can also improve the follower mechanism, where we can create more complex multi-topic link updates, where agreement on one topic, might lead to a following that is across all similar topics – reflecting closer to the way social connections form. This can be expanded to cover similar or shared geographies, hobbies and, lifestyles. Currently, we are limited by our real data sets on topics. We are in particular interested in any ability to model social acceptance of future emerging technologies (e.g., 6G), where there are no real data sets, but understanding social opinion is critical. We do note that there are emerging synthetic world simulations (e.g., academic \cite{park2023social} and commercial Project OdySSEy)  to emulate a realistic life and social interactions.

\section{Code \& Data Availability Statement}
The source code (Python) for running the simulations is published at https://github.com/AlminaJin/
InfluentialAgent. All data is available from public data bases. 

\section{Acknowledgments}
The work is supported by: (1) "Networked Social Influence and Acceptance in a New Age of Crises" funded by USAF OFSR (FA8655-20-1-7031); and (2) "Communications Hub For Empowering Distributed ClouD Computing Applications And Research" funded by EPSRC/DSIT (EP/X040518/1) (EP/Y037421/1).










\endparano



\section{Appendix A: GPT API Setting}
\begin{table}[ht]
\centering
\begin{tabular}{ccc}
\hline
        & Creative & Narrow\\
    \hline
    max\_tokens & 2000& 2000\\
    temperature & 1.4& 0.1\\
    top\_p & 0.95& 0.5\\
    n & 1& 1\\
    stop & "\}"& "\}"\\
    model & "gpt-3.5-turbo"& "gpt-3.5-turbo"\\
\hline
\end{tabular}
\caption{GPT Creative Parameters}
\label{Creative}
\end{table}

Table \ref{Creative} outlines the parameters set for the GPT API, distinguishing between 'creative' case and 'narrow' case configurations. These parameters are used in LLMs to tune how much creativity they add to the natural language output. The role of LLM is only to map the RL opinion decision to a natural language format. Therefore, this creativity is only in the mapping stage, not in the decision stage. Most literature \cite{park2022social} use these parameters similarly to map the boundaries of the LLM performance. Hopefully this explains that this does not affect how opinion influence decisions are made, but only the text format in which it manifests. Within these settings, higher temperature and top-p values are indicative of a setup geared towards fostering creativity. For tasks related to post generation and opinion elicitation, we employ the pre-trained GPT-3.5 model developed by OpenAI \cite{openai2022gpt35}.

\subsection{Appendix B: Opinion Example}
In this section, we will explore illustrative examples of 'strong pro' and 'strong con' concerning the concept of a 'society with no gender.' These examples are meant to elucidate the polarized viewpoints on this topic.

A 'strong pro' argument is presented as follows:
\begin{quote}
    "Any violent behavior targeted at the traits now associated with gender would be illogical, because that would assume that gender stereotypes would persist without gender. Such an assumption would almost certainly also include the academically false assumption that gender is biological."
\end{quote}

Conversely, a 'strong con' perspective is showed in the following statement:
\begin{quote}
    "It won't disappear. The idea of Sexist Violence in the eyes of society might, but that's only because what we call Gender-based violence is just that... Violence. What you would see in a drop in Gender based violence, there would be a spike in generalized Violence to take its place. Its only taking a tag of something and putting it into a generalized pool of everything else. It helps no one and Dis-benefits everybody involved. So a society without gender is a really bad idea."
\end{quote}

\subsection{Appendix C: Opinion Category Setting}

Based on the polarity distributions, we categorize the posts into distinct opinion groups while ensuring an even distribution across these categories. The specific range settings and the categorization results are presented below.

\begin{table}[ht]
\centering
\begin{tabular}{lll}
    Category & Polarity score & Number of posts\\ 
    \hline
   Strong pro & 1 -- 0.64 & 83\\
    Pro & 0.58--0.64 & 89\\
    Neutral & 0.51--0.58 & 128\\
    Con & 0.45--0.51 & 93\\
    Strong con & 0 -- 0.45 & 93\\
\end{tabular}
\caption{Gender opinion category setting}
\label{category}
\end{table}
\begin{table}[ht]
\centering
\begin{tabular}{lll}
    Category & Polarity score & Number of posts\\ 
    \hline
   Strong pro & 1 -- 0.55 & 101\\
    Pro & 0.48--0.55 & 128\\
    Neutral & 0.42--0.48 & 145\\
    Con & 0.37--0.42 & 102\\
    Strong con & 0 -- 0.37 & 184\\
\end{tabular}
\caption{Drug opinion category setting}
\label{Dcategory}
\end{table}

\subsection{Appendix D: Model Documentation}

The source code (Python) for running the simulations is published at \url{https://github.com/AlminaJin/InfluentialAgent.}





\qquad

\bibliographystyle{jasss}
\bibliography{bio} 


\end{document}